\documentclass[12pt]{article}
\usepackage[utf8]{inputenc}
\usepackage{amsmath}
\usepackage{graphicx,psfrag,epsf}
\usepackage{enumerate}
\usepackage{natbib}
\usepackage{url}

\newcommand{\blind}{0}

\usepackage{float}
\usepackage{amsthm,amssymb}
\usepackage{amsfonts}
\usepackage{mathrsfs}
\usepackage{float}
\usepackage{subfigure}
\setcitestyle{notesep={; },round,aysep={},yysep={;}}
\usepackage{hyperref}
\hypersetup{
	colorlinks=true,
	linkcolor=cyan,
	filecolor=black,
	urlcolor=red,
	citecolor=black,
}
\usepackage{amsbsy}
\usepackage{amsmath}
\usepackage{amssymb}
\usepackage{multirow}
\usepackage{graphicx}
\usepackage{varwidth}
\usepackage{bm}
\usepackage{color}
\usepackage{hyperref}
\usepackage{rotating}
\usepackage{graphicx}
\usepackage{epsfig}
\usepackage{epstopdf}
\usepackage{appendix}
\usepackage{algorithm}
\usepackage{algorithmic}
\usepackage{ulem}

\newtheorem{thm}{Theorem}

\newtheorem{corollary}{Corollary}

\newcommand{\btheta}{\bm{\theta}}
\newcommand{\bbeta}{\bm{\beta}}
\newcommand{\bgamma}{\bm{\gamma}}

\newcommand{\bx}{\bm{x}}

\newcommand{\ba}{\bm{a}}

\newcommand{\bW}{\bm{W}}
\newcommand{\bX}{\bm{X}}

\newcommand{\mcB}{\mathcal{B}}

\newcommand{\mcN}{\mathcal{N}}
\newcommand{\mcR}{\mathcal{R}}
\newcommand{\mcS}{\mathcal{S}}

\newcommand{\mcK}{\mathcal{K}}
\newcommand{\mcG}{\mathcal{G}}

\newcommand{\mcD}{\mathcal{D}}
\newcommand{\mcO}{\mathcal{O}}

\newcommand{\mbE}{\mathbb{E}}

\date{}

\begin{document}

\begin{sloppypar}
\def\spacingset#1{\renewcommand{\baselinestretch}%
{#1}\small\normalsize} \spacingset{1}


\if0\blind
{
  \title{\bf Deep regression learning  with optimal loss function}
  \author{Xuancheng Wang*, Ling Zhou\thanks{Co-first authors.} \  and  Huazhen Lin\thanks{Corresponding author. Email address: \emph{linhz@swufe.edu.cn}. } \thanks{The research was supported by National Key R\&D Program of China (No.2022YFA1003702), National Natural Science Foundation of China (Nos. 11931014 and 12271441), and  New Cornerstone Science Foundation.}\vspace{.5cm}\\
    New Cornerstone Science Laboratory, \\
    Center of Statistical Research and School of Statistics,\\
     Southwestern University of Finance and Economics, Chengdu, China}
  \maketitle
} \fi

\if1\blind
{
  \bigskip
  \bigskip
  \bigskip
  \begin{center}
    {\LARGE\bf Deep regression learning  with optimal loss function}
\end{center}
  \medskip
} \fi

\bigskip
\begin{abstract}
Due to powerful function fitting ability and  effective  training algorithms of neural networks, in this paper,  we develop a novel efficient and robust nonparametric regression estimator  under a framework of feedforward neural network (FNN).   There are several interesting characteristics for the proposed estimator. First, the loss function is built upon an estimated maximum likelihood function, who integrates the information from  observed data, as well as the information from   data structure. 
Consequently, the resulting estimator has desirable optimal properties, such as efficiency. 
Second,  different from the traditional maximum likelihood estimation (MLE), we do not require the specification of the distribution, hence the proposed  estimator  is  flexible  to  any kind of distribution, such as  heavy tails, multimodal or heterogeneous distribution.  Third, the proposed loss function relies on probabilities rather than direct observations as in least square loss, hence contributes the robustness in the proposed estimator. Finally, the proposed loss function involves nonparametric regression function only. This enables the direct application of the existing packages, 
and thus the computation and programming are simple.
We establish the large sample property of the proposed estimator in terms of its excess risk and  minimax near-optimal rate. The theoretical results demonstrate that 
the proposed estimator is equivalent to the true MLE in which the  density function is known. Our simulation studies show that  the proposed estimator outperforms the existing methods in terms of prediction accuracy, efficiency and robustness. Particularly, it is comparable to  the true MLE, and  even gets  better  as the sample size increases.  This implies that the adaptive and data-driven loss function from  the estimated density  may  offer an additional avenue for capturing valuable information.  
We further apply the proposed  method  to four real data examples, resulting in significantly reduced out-of-sample prediction errors compared to existing methods.
\end{abstract}

\noindent%
{\it Keywords:}  Estimated maximum likelihood estimation, feedforward neural network, excess risk, kernel density estimation.
\vfill

\newpage
\spacingset{1.45} 
\section{Introduction}
\label{sec:intro}

Consider a nonparametric regression model,
\begin{eqnarray}
\label{model}
		Y&=&g(\bX)+\epsilon,
	\end{eqnarray}
where $Y \in \mathbb{R}$ is a response variable, $\bX \in \mathcal{X} \subseteqq \mathbb{R}^{d} $ is a $d$-dimensional vector of predictors, $g:\mathcal{X} \rightarrow \mathbb{R} $ is an unknown regression function, $\epsilon$ is an error independent of $\bX$. Nonparametric regression is a  basic and core problem in statistics and machine learning, where the  purpose is  estimating the unknown target regression function $g$ given independent and identically distributed (i.i.d.) samples $S\equiv {(\bX_{i},Y_{i})}_{i=1}^{n}$ with  the sample size  $n$.  Since the distribution of $\epsilon$ is unknown,  $g(\cdot)$  is usually estimated based on the least square (LS) criterion, that is,
\begin{eqnarray}
\label{ls}
\hat g=\underset{g: \mathbb{R}^{d}\rightarrow \mathbb{R}}{\arg \min} \frac{1}{n}\sum_{i=1}^n\left\{Y_i-g(\bX_i)\right\}^2.
\end{eqnarray}
    Driven by  various nonparametric approximation techniques,  there is a vast literature  on nonparametric regression. For example,  tree regression \citep{breiman2017classification}, random forests \citep{breiman2001random},  and nonparametric smoothing methods such as nearest neighbor regression  \citep{cheng1984strong, devroye1994strong},  kernel regression \citep{nadaraya1964estimating, watson1964smooth, hall2001nonparametric}, local polynomial regression \citep{ fan2018local}, spline approximation \citep{schumaker2007spline} and reproducing kernel regression \citep{berlinet2011reproducing,lv2018oracle}, among others.

Recently, attributed to powerful function fitting ability, well-designed neural network architectures and  effective  training algorithms and
high-performance computing technologies,  deep neural network (DNN) with the empirical LS loss function  has enjoyed tremendous success in a variety of applications, 
such as the fields of computer vision, 
natural language processing, 
speech recognition, 
among others.   Based on the theoretical results concerning approximation error and stochastic error,  with the LS loss,  several inspiring works  have obtained   the minimax near-optimal rate at  $n^{-\frac{2\beta}{2\beta+d}}(\log n)^{s}$  for   learning the regression function $g$  under feedforward neural network (FNN),  
 with the assumption that  $g$  is $\beta$-H$\ddot o$lder smooth. 
In these works, the response variable or the error term  is assumed to be bounded \citep{gyorfi2002distribution,farrell2021deep}, have finite $p$-th moment with $p > 1$ \citep{kohler2021rate,kohler2022estimation}, sub-Gaussian \citep{bauer2019deep, chen2019nonparametric, schmidt2019deep,  schmidt2020nonparametric, fan2022factor,bhattacharya2023deep}, sub-exponential \citep{jiao2021deep,yan2023nonparametric} or have finite variance \citep{liu2022optimal}.


The LS criterion based estimators are mathematically convenient,  easily implemented, and  efficient when the error $\epsilon$ is normally distributed. However, as it is expressed in (\ref{ls}), the LS loss is sensitive to large errors, that is, the LS estimator is  severely influenced by outliers, resulting in unstable and unreliable estimation. In the era of ``big data",
data generation mechanism and collection are unmanageable, and thus non-Gaussian noises or outliers are  almost inevitable. 
To address the unstableness,  a lot of robust methods based on traditional nonparametric regression techniques have been developed,  for example, the kernel M-smoother \citep{hardle1989asymptotic}, median smoothing\citep{tukey1977exploratory} , locally weighted regression \citep{stone1977consistent,cleveland1979robust},  the local least absolute method \citep{wang19941}, quantile regression \citep{koenker1978regression,he2013quantile,lv2018oracle}, among others.

Recently, within the framework of FNN, several robust methods have been introduced to address non-Gaussian noise problems, and corresponding convergence rates for learning the function $g$ have also been established.
For instance, \citet{lederer2020risk,shen2021robust} and \cite{fan2022noise}  have explored non-asymptotic error bounds of the estimators that minimizing  robust loss functions, 
such as the least-absolute deviation loss \citep{bassett1978asymptotic}, Huber loss \citep{huber1973robust},  Cauchy loss and Tukey's biweight loss \citep{beaton1974fitting}. 
Particularly, based on  a general robust loss function satisfying a Lipschitz continuity, \citet{farrell2021deep} have demonstrated the  convergence rate $n^{-\frac{2\beta}{2\beta+d}}(\log n)^{4}$   with the assumption that the response is bounded, which means that heavy-tail error is not applicable.
To relax the bounded restriction on the response, \citet{shen2021robust} and \citet{shen2021deep} have established   the convergence rate $n^{-\frac{2\beta}{2\beta+d}+ 1/p}(\log n)^{c}$  under the assumption that the $p$-th moment of response is bounded for some $p>1$. 
These methods are proposed for improving the robustness, 
they are sub-optimal  in terms of efficiency.  


This work attempts to  provide a loss function  which is   efficient as well as  robust for nonparametric regression estimation within the framework of FNN.  
It is worth noting that in the least squares (LS) criterion, observations in which the response variable $Y_i$ deviates significantly from the conditional mean $g(\bX_i)$ play a significant  role, which may seem counterintuitive. In fact, when estimating the conditional mean $g(\cdot)$, observations in which $Y_i$ is closer to $g(\bX_i)$ are supposed to logically carry more information than those where the response is away from the conditional mean. This can be expressed in terms of probability of observation.
Therefore, we
 propose a loss function  based on the estimated likelihood  function,  which has the form of
\begin{eqnarray}
\label{mle}
\hat g=\underset{g: \mathbb{R}^{d}\rightarrow \mathbb{R}}{\arg \max} \frac{1}{n}\sum_{i=1}^n\log \hat{f}(Y_i-g(\bX_i)),
\end{eqnarray}
where $\hat{f}$ is an estimator of the density function of $\epsilon_i=Y_i-g(\bX_i)$.  
We simplify the  FNN estimators of $g(\cdot)$ based on 
maximizing  Estimated log-Likelihood functions (\ref{mle}) by EML-FNN, which is expected to have the desirable optimal properties since
we  use the density function and leverage the data distribution. In addition, different from the traditional maximum
likelihood estimator (MLE) where $f(\cdot)$ is known, the proposed EML-FNN  is flexible  as it  avoids  specifying the error distribution.
Moreover, the quasi-likelihood loss (\ref{mle}), which relies on probabilities rather than direct observations as in LSE, contributes the robustness of the proposed EML-FNN.
More interesting, in comparison to the MLE where $f(\cdot)$ is known, the adaptive form via estimating the density $f(\cdot)$ in (\ref{mle}) proves to be  effective in learning  the data structure and  offers an additional avenue for capturing  information. This  is supported by our simulation studies.  Specifically, Figures \ref{Fig.1} to \ref{Fig.3} reveal the following results: 
when $\varepsilon_i$ follows a normal distribution where the LSE is  equivalent to the MLE,  the EML-FNN performs slightly  better than FNN estimators based on Least Square Error (LSE-FNN) for data with a  larger sample size ($n=1024$).   
However,  when $\varepsilon_i$ deviates  from a normal distribution or has heterogeneous variances, the EML-FNN significantly outperforms the LSE-FNN. The enhanced performance may be attributed to the utilization of structural information  via  the estimated density function $\hat f(\cdot)$.

With the explicit form of Nadaraya–Watson kernel estimator for the density function of $\epsilon_i$, 
we develop a FNN estimation  for $g$   that circumvents  estimating the unknown density function, 
resulting in an objective function that solely involves $g$. This enables
the direct application of existing packages Pytorch \citep{paszke2019pytorch} and Scikit-learn \citep{scikit-learn} in python, simplifying the computation and programming.
We establish the large sample property of $\hat g$ in terms of its excess risk  and the minimax rate, which demonstrate that the proposed estimator for $g$ is equivalent to the one based on (\ref{mle}) when the  density function is known. As a result, the proposed deep learning approach for $g$ exhibits the desired optimal properties, such as efficiency \citep{zhou2018efficient,zhou2019efficient}.
Finally, we employ the proposed method to analyze four real  datasets.  
Table \ref{table:1} shows that the proposed EML-FNN   provides much higher prediction accuracy than the existing  methods for each dataset.

The paper is structured as follows. In Section 2, we  introduce the proposed EML-FNN. 
In Section 3, we establish the large sample property of $\hat g$ in terms of its excess risk  and the minimax rate. Section 4 provides simulation studies to investigate the performance of  the proposed method via the comparison with the competing estimation methods.
In Section 5, we apply the proposed method to analyze four real data. We conclude the paper with a discussion in Section 6. Technical proofs are included in the Supplementary Material.

\section{Method}
\label{sec:meth}


We estimate $g$ under the framework of FNN. In particular, we set  $\mcG$ to be a function class consisting of ReLU neural networks, that is,
$\mcG := \mcG_{\mathcal{D},\mathcal{U},\mathcal{W},\mathcal{S},\mathcal{B}},$
where the input data is the predictor  $X$, forming the first layer, and the output is the last layer of the network; Such a network $\mathcal{G}$ has $\mathcal{D}$ hidden layers and a total of $(\mathcal{D} + 2)$ layers. Denote the width of  layer $j$ by   $d_j$, $j=0, \cdots, \mcD, \mcD+1$ with  $d_0 = d$ representing  the dimension of the input $X$, and $d_{\mathcal{D}+1} = 1$ representing  the dimension of the response $Y$.
The width $\mathcal{W}$ is defined as the maximum width among the hidden layers, i.e., $\mathcal{W} = \max{(d_1, ..., d_{\mcD})}$. The size $\mathcal{S}$ is defined as the total number of parameters in the network $\mathcal{G}$, given by $\mathcal{S} = \sum_{i=0}^{\mathcal{D}} d_{i+1} \times (d_{i} + 1)$; The number of neurons $\mathcal{U}$ is defined as the total number of computational units in the hidden layers, given by $\mathcal{U} = \sum_{i=1}^{\mathcal{D}} d_{i}$.  Further, we assume every function $g \in \mathcal{G}$ satisfies $| g |_{\infty} \leq \mathcal{B}$ with $\mathcal{B}$ being a positive constant.

With $g\in \mcG$, $g$ can be estimated by
\begin{eqnarray}
\label{obj0}
 \arg\min_{g\in \mcG} \left\{ \frac{1}{n}\sum_{i=1}^n  \rho(Y_i - g(\bX_i))\right\},
\end{eqnarray}
where $\rho(\cdot)$ is a given loss function, for example, the least squares $\rho(t)=t^2$; least absolute criteria  $\rho(t)=|t|$; Huber loss,  Cauchy loss,  and Tukey’s biweight loss, and so on.
The LS based estimator is efficient only when the error $\epsilon$ is normally distributed. 
The estimators based on robust loss functions such as least absolute, Huber, Cauchy and Tukey’s biweight are   robust  but they are sub-optimal in terms of efficiency.

When $f(\cdot)$ is known, an ideal estimator of $g$ can be obtained by, 
\begin{eqnarray}
\label{obj1}
\hat{g} = \arg\min_{g\in \mcG} \mcR_n(g) := \arg\min_{g\in \mcG} \left\{ \frac{1}{n}\sum_{i=1}^n \left( -\log f(Y_i - g(\bX_i)) \right)\right\}.
\end{eqnarray}
However, in reality, $f$ is usually unknown.  To ensure that we do not misspecify  the distribution and simultaneously obtain an estimator based on optimal loss, we employ kernel techniques to estimate the density function  $f$. 
That is,
\begin{eqnarray}
\label{ff}
\hat{f}(z) = \frac{1}{n}\sum_{i=1}^n\mcK_h(\epsilon_i, z),
\end{eqnarray}
where $\mcK_h(y_1, y_2) = K(\frac{y_1 - y_2}{h})/h$, $h$ is a bandwidth and $K(\cdot)$ is a  kernel function.


Replacing $f(\cdot)$ in (\ref{obj1}) with $\hat f$, we estimate $g$ by, 
\begin{eqnarray}
\label{eq:me}
\hat{g} = \arg\min_{g \in \mcG}\hat\mcR_n(g) = \arg\min_{g \in \mcG}  n^{-1}\sum_{i=1}^n \left( -\log \hat{f}(Y_i - g(\bX_i)) \right).
\end{eqnarray}
That is,
\begin{eqnarray}
\label{eq:me1}
\hat{g} = \arg\min_{g \in \mcG}\hat\mcR_n(g)= \arg\min_{g \in \mcG}  n^{-1}\sum_{i=1}^n \left( -\log  \frac{1}{n}\sum_{j=1}^n\mcK_h(Y_j - g(\bX_j), Y_i - g(\bX_i)) \right).
\end{eqnarray}

Recall that the conventional FNN \citep{chen2019nonparametric, nakada2019adaptive, schmidt2020nonparametric, kohler2021rate, jiao2021deep, kohler2022estimation, liu2022optimal,fan2022factor,yan2023nonparametric,bhattacharya2023deep} minimizes a least square objective and   is sensitive to the data's distribution type and outliers, which leading to the development of the robust FNN \citep{lederer2020risk,shen2021robust,fan2022noise}. However, 
the enhanced  robustness  comes at the cost of efficiency. 
In contrast to existing methods, our approach stands out by utilizing a MLE criterion as the objective function, thereby achieving both efficiency and robustness. In particular, efficiency is attained by fully leveraging  the data distribution,  robustness is added to our estimator because  our proposed loss function relies on probabilities rather than direct observations as in LS.
Moreover, the kernel-based estimation approach benefits from the smooth continuity of kernel functions, facilitating gradient calculations and overcoming non-differentiability issues  when dealing with   densities such as  the uniform distribution, mixture distribution, and heteroscedasticity. Finally,  the proposed loss function (\ref{eq:me1}) involves $g$ only. This enables the direct application of packages   Pytorch \citep{paszke2019pytorch} and Scikit-learn \citep{scikit-learn} in python, simplifying the computation and programming.

The proposed $\hat{\mcR}_n(g)$ involves a tuning parameter $h$. According to the property of kernel approximation, a smaller $h$ yields a more accurate density approximation but with a larger variance. Fortunately, the summation over individuals mitigates the increased variance caused by a small $h$. 
Therefore, when computational feasibility allows, a smaller value of $h$ is preferred.
The conclusion is supported by both our theoretical and numerical results. 
In practice,  we use the Gaussian kernel function and set $\hat{f} = 1e - 5$ when $\hat{f} < 1e - 5$ because logarithmic transformation is required in the objective function.


\section{Large sample properties}
\label{theory}
In this section, we establish the large sample property of $\hat{g}$ in terms of its excess risk, 
which is defined as the difference between the risks of ${g}$ and $g^*$:
\[
\mcR(g) - \mcR(g^*) = \mbE \left( - \log f (Y_i - g(\bX_i)) \right) -  \mbE \left( - \log f (Y_i - g^*(\bX_i)) \right),
\]
where $g^*$ is defined as
\[
g^* := \arg\min_{g} \mcR(g) = \arg\min_{g}\mbE \left( - \log f\left(Y_i - g(\bX_i)\right)\right). 
\]
The minimizer is taken over the entire space, and thus implies that $g^*$ does not necessarily belong to the set $\mathcal{G}$. We further define $g_{\mcG}^* := \arg\min_{g\in \mcG} \mcR(g)$ in the set $\mathcal{G}$.

Denote $f^{(r)}(\cdot)$ to be  the $r$th derivative of $f$, and $f_{\bx}(\cdot)$ to be the density function of covariates $\bX$,  who is supported on a bounded set, and for simplicity, we assume this bounded set to be $[0, 1]^d$.  In the rest of the paper, the symbol $c$ denotes a positive constant which may vary across different contexts.  The following conditions are required  for establishing  the rate of the excess risk:
\begin{enumerate}
\item[{(C1)}]  Kernel:  Let $U_r = \int K(t) t^r dt$ and $v_r = \int K^2(t)t^r dt $. Assume the kernel function $K(\cdot)$ has a bounded second derivative, $U_0 = 1$ and $U_1 = 0$.\label{C1}
\item[{(C2)}] Bandwidth: $h \to 0$ and  $nh \to \infty$ as $n\rightarrow\infty$.\label{C2}
\item[{(C3)}]  Density function $f$:  Assume the density function $f(\cdot)$ has a continuous second derivative and satisfies $f(\epsilon) > c > 0$ for any $\epsilon$ belonging to the support set of $f$.
\label{C3}
\item[{(C4)}] Function class for $g$ and $g^*$: For any function $g \in \mcG$ and the true function $g^*$, we assume $\|g\|_{\infty}< \mcB$ and $\|g^*\|_{\infty} < \mcB$. 
\label{C4}
\end{enumerate}

Condition~(C1) is a mild condition for the kernel function, which is easy to be satisfied when the kernel function is a symmetric density function. 
Condition~(C2) is  the most commonly used assumption  for the bandwidth. Condition (C3) requires  a lower bound of the density function  to avoid  tail-related problems. 
The simulation studies, where the lower bounded condition is not met  across all four distributions, demonstrate that the proposed estimator maintains its effectiveness even in scenarios where the condition does not hold. 
Condition~(C4) is a bounded condition for the function class $\mcG$ and the true function $g^*$, which is commonly used in \citet{shen2019deep,lu2021deep,chen2019nonparametric,yarotsky2017error}. 
It is noteworthy that, in cases where the explicit depiction of the approximation error for the function class $\mathcal{G}$ to  $g^*$ becomes necessary, an additional condition will be introduced concerning the category to which $g^*$ belongs. This is demonstrated in Corollary~\ref{cor:rate}.

Define  $\mcG|_{\bx} := \{g(\bx_1), g(\bx_2), \cdots, g(\bx_n): g \in \mcG\}$ for a given sequence $\bx = (\bx_1, \cdots, \bx_n)$ and denote  $\mcN_{2n}(\delta, \|\cdot\|_{\infty}, \mcG|_{\bx})$ to be the covering number of $\mcG|_{\bx}$ under the norm $\|\cdot\|_{\infty}$ with radius $\delta$. Let $A \preceq B$ represent $A \leq c B$ for a postive constant $c$. In the following Theorems~\ref{th:oracle} and \ref{th:error-one},  we  show  the excess risk of the proposed estimator  under the true density function and the estimated density function to see how much  difference of the proposed estimator  from the true MLE estimator, which is defined as
\begin{eqnarray}
\label{eq:oracle}
\hat{g}_{oracle} = \arg\min_{g \in \mcG} \left\{
\frac{1}{n}\sum_{i=1}^n \big( - \log f(Y_i - g(\bX_i)) \big)
\right\}.
\end{eqnarray}

\begin{thm}
\label{th:oracle}
Under Conditions (C3) and (C4),  we have that, as $n\rightarrow\infty$,
\[
\mbE\left(
\mcR(\hat{g}_{oracle}) - \mcR(g^*) \right) \preceq \frac{\log \mcN_{2n}(n^{-1}, \|\cdot\|_{\infty}, \mcG|_{\bx})}{n} + \big(\mcR(g^*_{\mcG}) - \mcR(g^*)\big).
\]
\end{thm}

Recall that  the excess risk of the LS estimator, takes the form: $\frac{ \mathcal{B}^2 \log 2\mcN_{2n}(n^{-1}, |\cdot |_{\infty}, \mcG|_{\bx})(\log n)^{c}}{n} + \left(\mcR(g^*_{\mcG}) - \mcR(g^*)\right)$  for some positive constant $c$ with the condition of  bounded response  \citep{gyorfi2002distribution,farrell2021deep} or bounded $p$-th moment \citep{kohler2021rate,kohler2022estimation}. For the robust loss  considered in \citet{shen2021robust}, the excess risk has the form of: $\frac{ \lambda_{L} \mathcal{B} \log 2\mcN_{2n}(n^{-1}, |\cdot|_{\infty}, \mcG|_{\bx})(\log n)^c}{n^{1-1/p}} + \left(\mcR(g^*_{\mcG}) - \mcR(g^*)\right)$,  where $p$ represents the bounded  $p$-th moment of the outcome, and 
$\lambda_{L}$ represents the Lipschitz coefficient of robust loss function. 
Clearly, the oracle  estimator $\hat{g}_{oracle}$ presents a slightly more favorable excess risk bound compared to the OLS estimator,  as it lacks the $(\log n)^c$ multiplier. 
Additionally, our estimator converges faster than the robust estimators with a rate of $(\log n)^c/n^{1-1/p}$ for robust estimators versus a reduced rate of $1/n$ for our estimator in  estimation error.
It is important to highlight that, unlike the requirement of a Lipschitz condition for the robust loss, we instead invoke a lower bound   condition~(C3) for the density function. The introduction of a lower bound to the density function  is helpful to the stability of our estimator. 
On the other hand,  by leveraging the inherent benefits of the density function, our proposed estimator exemplifies a harmonious blend of robustness and efficiency that is crucial for practical applications.

\begin{thm}
\label{th:error-one}
For  the proposed estimator $\hat g$, under conditions (C1)--(C4), we have
\begin{eqnarray*}
\mbE\left(
\mcR(\hat{g}) - \mcR(g^*) \right) &\preceq&
\left( \frac{ \log \mcN_{2n}(\frac{1}{n}, \|\cdot\|_{\infty}, \mcG|_{\bx})}{n} \right)
  + \left(\mcR(g^*_{\mcG}) - \mcR(g^*)\right)+\left( \|g^*_{\mcG} - g^*\|^2_{\infty} + h^2\right).
\end{eqnarray*}
\end{thm}

Theorems \ref{th:oracle} and \ref{th:error-one} shows that the upper bounds of the excess risk for both $\hat{g}_{oracle}$ and $\hat{g}$ encompass two terms: $\frac{\log \mathcal{N}{2n}(\frac{1}{n}, |\cdot|{\infty}, \mathcal{G}|{\mathbf{x}})}{n}$ and $\mathcal{R}(g^*_{\mathcal{G}}) - \mathcal{R}(g^*)$, which  represent the estimation error of $\hat{g}$  evaluated at the true density function $f$, and the approximate bias of the FNN space towards the true  function $g^*$, respectively. The disparity in excess risks between $\hat{g}_{oracle}$ and $\hat{g}$ is encapsulated in $\|g^*_{\mathcal{G}} - g^*\|^2_{\infty} + h^2$, which describes the  error introduced by substituting $f$ with its kernel estimator $\hat{f}$.  The  error  implies that utilizing the kernel estimator $\hat{f}$ in lieu of $f$ does not introduce additional variance. However, it does lead to significant approximation bias  when using a larger value of $h$, thus advocating the preference for a smaller value of $h$ to mitigate this bias, particularly, the bias 
is ignorable if   $h^2 \preceq  \frac{ \log \mcN_{2n}(\frac{1}{n}, \|\cdot\|_{\infty}, \mcG|_{\bx})}{n}$  and the FNN function closely approximates the true function $g^*$. The former can be satisfied by taking a small $h$ and the later holds due to powerful function fitting ability of FNN. 
The simulation studies in Section \ref{sec:ss} further confirm the conclusion.

With the discussion above,  we hence investigate the efficiency of the proposed estimator via that for 
the oracle estimator $\hat{g}_{oracle}$.  
For simplicity, we assume $g^* = g^*_{\mcG}$, that is, the true function belongs to the FNN space. Recall that for  $g \in \mcG$, we have
\[
g(x) = \bW^\top_{\mcD}\sigma\left(\bW^\top_{\mcD-1}\sigma(\bW^\top_{\mcD-2}\sigma(\bW_{\mcD-3} \cdots \sigma(\bW^\top_{0}\bX + \ba_0)) + \ba_{\mcD-2}) + \ba_{\mcD-1}\right) + a_{\mcD},
\]
where $\sigma(\cdot)$ is a given activation function and $\bW_{r}, \ba_{r}, r = 0, \cdots, \mcD$ are parameters. Then, we can write  $g(\bx) = g(\bx; \btheta)$  with  $\btheta=\{\bW_{r}, \ba_{r}, r = 0, \cdots, \mcD\}$. Denote   $g^*(\bx) = g(\bx; \btheta^*)$.  
We can obtain that  
$\hat{g}_{oracle}(\bx) = g(\bx; \hat{\btheta}_{o})$ with $\hat{\btheta}_{o}$ satisfying $\hat{\btheta}_o = \arg\min_{\btheta} \left\{
\frac{1}{n}\sum_{i=1}^n \big( - \log f(Y_i - g(\bX_i; \btheta)) \big)
\right\}.$ If $\mbE\left[\left( \frac{d \log f(Y - g(\bX; \btheta))}{d \btheta} \right)\left( \frac{d \log f(Y - g(\bX; \btheta))}{d \btheta} \right)^\top\right]$ is positive definite around $\btheta^*$ and $\mbE(\hat{\btheta}_o) = \btheta^*$,   we have $\mbE(\hat{\btheta}_o\hat{\btheta}_o^\top) = n^{-1}\left\{\mbE\left[\left( \frac{d \log f(Y - g(\bX; \btheta))}{d \btheta} \right)\left( \frac{d \log f(Y - g(\bX; \btheta))}{d \btheta} \right)^\top\mid_{\btheta = \btheta^*}\right]\right\}^{-1}$ \citep{onzon2011multivariate}. Then for any unbiased estimator  $\check{\btheta}$ that  $\mbE(\check{\btheta}) = \btheta^*$, based on the Multivariate Cramer-Rao Lower Bound, it holds that
\[
\mbE(\check{\btheta}\check{\btheta}^\top) \succeq  n^{-1}\left\{\mbE\left[\left( \frac{d \log f(Y - g(\bX; \btheta))}{d \btheta} \right)\left( \frac{d \log f(Y - g(\bX; \btheta))}{d \btheta} \right)^\top\mid_{\btheta = \btheta_o}\right]\right\}^{-1} = \mbE(\hat{\btheta}_o\hat{\btheta}_o^\top).
\]
which leads that $\text{Var}(\check{\btheta}) \succeq \text{Var}(\hat{\btheta}_o)$, where $A \succeq B$ represents $A - B$ is a semi-positive matrix.  Combining with the delta method, it holds that
$\text{Var}(\check{g}) := \text{Var}(g(\bx; \check{\btheta})) \geq \text{Var}(\hat{g}_{oracle})$.  From this perspective, we can  characterize $\hat{g}_{oracle}$ as an efficient estimator, while $\hat{g}$ also possesses such efficiency under certain straightforward conditions, such as  $h^2 \preceq \frac{\mcS}{n}$, where $\mcS$ is the length of $\btheta$.



Now, we further explore how the excess risk relies on FNN structure, as well as  the function class  which  $g^*$ belongs to. Let $\beta = s + r$, $r\in (0,1]$ and $s = \left \lfloor \beta \right \rfloor \in \mathbb{N}_0$, where $\left \lfloor \beta \right \rfloor$ denotes the largest integer strictly smaller than $\beta$ and $\mathbb{N}_0$ denotes the set of non-negative integers. For a finite constant $B_0>0$, the $H\ddot o lder$ class $\mathcal{H}_\beta([0,1]^d,B_0)$ is defined as
\begin{eqnarray*}
		\mathcal{H}_\beta([0,1]^d,B_0)=\{g:[0,1]^d \mapsto \mathbb{R},\max_{\| \alpha \|_1 < s} \| \partial^{\alpha} g \|_{\infty} \leq B_0, \max_{\| \alpha \|_1 = s} \sup_{x \neq y} \frac{\left | \partial^{\alpha} g(x) - \partial^{\alpha} g(y) \right |}{\| x-y \|^r_2} \leq B_0       \}
	\end{eqnarray*}
where $\partial^\alpha = \partial^{\alpha_1} \cdots \partial^{\alpha_d}$ with $\alpha=(\alpha_1,\cdots,\alpha_d)^{T} \in \mathbb{N}_0^d$ and $\| \alpha \|_1 = \sum_{i=1}^{d}\alpha_i$.

Denote $\left \lceil a \right \rceil$ to be  the smallest integer no less than $a$ and  $\mathbb{N}^+$ to be the set of positive  integers. Based on Lemma 1 of \citet{jiao2021deep} for the approximation error in terms of FNN structures  and Lemma 2 of \citet{bartlett2019nearly} for the bounding covering number, we can conclude 
the following Corollary~\ref{cor:rate} from Theorem~\ref{th:error-one}:

\begin{corollary}
\label{cor:rate}
Given $H\ddot o lder$ smooth functions $g^* \in \mathcal{H}_\beta([0,1]^d,B_0)$, for any $D \in \mathbb{N}^+$, $W \in \mathbb{N}^+$, under conditions of Theorem~\ref{th:error-one}, Lemma 1 in \citet{jiao2021deep} and Lemma 2 in \citet{bartlett2019nearly}, if the FNN with a ReLU activation function has width $\mathcal{W} = C_3 (\left \lfloor \beta \right \rfloor +1)^{2}d^{\left \lfloor \beta \right \rfloor+1}W\left \lceil \log_2(8W) \right \rceil$ and depth $\mathcal{D} =C_4(\left \lfloor \beta \right \rfloor +1)^{2}D\left \lceil \log_2(8D) \right \rceil$, then
\begin{eqnarray*}
\mbE\left(
\mcR(\hat{g}) - \mcR(g^*) \right) &\preceq&
\frac{\mathcal{S}\mathcal{D} \log (\mathcal{S})  }{n}  +  h^2  + (WD)^{-4\beta/d} .
\end{eqnarray*}
\end{corollary}

In Corollary~\ref{cor:rate}, the first term comes  from the covering number of $\mcG$, which is bounded by its VC dimension
$\mcN_{2n}(\frac{1}{n}, |\cdot|_{\infty}, \mcG|_{\bx}) = O(\mcS \mcD \log (\mcS))$ \citep{bartlett2019nearly}, where $\mcS$ and $\mcD$ are the total number of parameters and hidden layers, respectively. The third term follows from the approximation results from  \citet{jiao2021deep} that
$\left \|g^* - g_{\mcG}^*\right \|_{\infty} \leq 18 B_0(\left \lfloor \beta \right \rfloor +1)^{2}d^{\left \lfloor \beta \right \rfloor+\max\{ \beta, 1 \}/2}(WD)^{-2\beta/d}$ and $\mbE(\mcR(g_{\mcG}^*) - \mcR(g^*)) \simeq \|g_{\mcG}^* - g^*\|^2_{\infty}$ where $A \simeq B$ represents $A \preceq B$ and $B \preceq A$.

If given $\mcS=\mcO(n^{\frac{d}{2\beta+d}}\log n)$ and $\mcD=\log n$, following Corollary~\ref{cor:rate} and $\mcS = O(\mathcal{W}^2\mcD)$, it holds that $\mbE\left(\mcR(\hat{g}) - \mcR(g^*) \right) \preceq n^{-\frac{2\beta}{2\beta+d}}(\log n)^3+ h^2  + n^{-\frac{2\beta}{2\beta + d}}$.  Hence, we have the following Corollary~\ref{optimal rate}.

\begin{corollary}
    \label{optimal rate}
Under the conditions in  Corollary~\ref{cor:rate}, if  $h^2 = O(n^{-\frac{2\beta}{2\beta + d}})$,  then
$$\mbE\left(\mcR(\hat{g}) - \mcR(g^*) \right) \preceq n^{-\frac{2\beta}{2\beta+d}}(\log n)^3,$$
which is comparable to $n^{-\frac{2\beta}{2\beta + d}}$,  the lower bound of the minimax learning rate of $g^*$ \citep{stone1982optimal},  i.e., $\min_{\check{g}}\max_{g^* \in \mathcal{H}_\beta([0,1]^d,B_0)}\mbE\left[\int_{[0, 1]^d} (\check{g}(\bx) - g^*(\bx))^2f_{\bx}(\bx)d\bx\right] \succeq n^{-\frac{2\beta}{2\beta + d}}$, where $\check{g}$ is an estimator of $g^*$  based on the data set $S$ and the expectation is taken with respect to the randomness of $S$.
\end{corollary}

It is interesting to compare several established convergence rates  under the framework of FNN. Particularly,  using the LS loss for $g\in \mathcal{H}_\beta([0,1]^d,B_0)$,  \cite{chen2019nonparametric,nakada2019adaptive,schmidt2020nonparametric,jiao2021deep,liu2022optimal,bhattacharya2023deep} and \cite{yan2023nonparametric} have obtained upper bound of the minimax learning rate of $g$ at  $n^{-\frac{2\beta}{2\beta+d}}(\log n)^{s}$,  which is nearly minimax optimal \citep{donoho1995wavelet,stone1982optimal}; Using a Lipschitz continuous loss function,  \citet{farrell2021deep} have obtained the  convergence rate at $n^{-\frac{2\beta}{2\beta+d}}(\log n)^{4}$  under a bounded response condition;  \citet{shen2021robust} and \citet{shen2021deep}  have obtained the convergence rate $n^{-\frac{2\beta}{2\beta+d}+ 1/p}(\log n)^{c}$ under the assumption of the bounded  $p$-th moment   for some $p>1$ to allow  heavy-tailed response $Y$. The severity of heavy tail decreases as $p$ increases.  In particular, if the response $Y$ is sub-exponentially distributed, $p = \infty$, the convergence rate achieves the minimax near-optimal rate. Obviously,  the proposed EML-FNN also enjoys nearly minimax optimal rate under condition (C3) with  the lower bounded restriction on the density function. It seems that, to achieve the optimal convergence rate, a bounded condition on the tail-probability may be essential. In fact,  a similar condition also appears for a quantile regression model,  that  under the assumption that the conditional density of $Y$ given $\bX$ around $\tau$-th quantile is bounded by below,  \citet{padilla2020quantile} have obtained the convergence rate $n^{-\frac{2\beta}{2\beta+d}}(\log n)^{2}$ for  a  quantile regression model under the framework of FNN.




\section{Simulation study}
\label{sec:ss}

We investigate the performance of the proposed EML-FNN (simplified as EML) by comparing it with the least square  estimator (LSE) and several robust methods with  $g$  approximated by FNN.  We consider four commonly used robust methods:
(1) Least absolute deviation method (LAD) with the loss function $\rho(x)=\left | x \right |$; (2) Huber method with the loss function  $\rho(x)=0.5 x^2I(\left | x \right |\leq \zeta)+(\zeta \left | x  \right | - \zeta^{2}/2)I(\left | x \right |> \zeta)$ at $\zeta = 1.345$;
(3)  Cauchy method with the loss function $\rho(x)=\log \{ 1+\kappa^{2}x^2 \}$ at $\kappa = 1$; (4)
Tukey’s biweight method with the loss function  $\rho(x)=t^2 [1-\{ 1-(x/t)^2 \}^3]I(\left |x\right |\leq t)/6+ t^2 I(\left |x\right |> t)/6$ at  $t = 4.685$.
We also investigate  the effect of bandwidth on our method in Section \ref{band}.

All feedforward network architecture, initial values, and data for training  are the same for all  methods involved. 
The computations were implemented  via packages  Pytorch \citep{paszke2019pytorch} and Scikit-learn \citep{scikit-learn} in python.
Specifically, we use  the network \textit{Net-d5-w256} with Relu activated functions, which  comprises  of 3 hidden layers, resulting in a network depth of 5 with  the corresponding network widths  $(d, 256, 256, 256, 1)$.    We use the Adam optimization algorithm  \citep{kingma2014adam}  with a learning rate of $0.0003$ for network parameters initialized by uniform distribution \citep{he2015delving}.  The coefficients used for calculating the running average of the gradient and the squared gradient are $\beta= (0.9,0.99)$.  We set the  training batch size to be equal to the size of the training data $n$, and train the network for at least 1000 epochs using a dropout rate of $0.01$  until the training loss converges or reaches a satisfactory level.  

To enhance flexibility and simplicity, we adopt a varying bandwidth  $h(\epsilon_i)=|\max(\epsilon_i(v))-\min(\epsilon_i(v))|,$  where the set $\epsilon_i(v)$ is  the neighborhood of  $\epsilon_i$ encompassing a proportion $v$ of the total sample \citep{loftsgaarden1965nonparametric}.  Then the selection of the bandwidth is translated into  selecting a value for $v$ from the interval $(0,1]$.
The constrained interval  simplifies the process of bandwidth selection.

We evaluate the performance of $\hat{g}$ by the  bias, standard deviation (SD) and root mean square error(RMSE), defined as
$    bias=\left [ \frac{1}{n_{grid}}\sum_{i=1}^{n_{grid}} (E\widehat{g}(z_{i})-g(z_{i}))^{2} \right ]^{\frac{1}{2}}$,
$	SD=\left [ \frac{1}{n_{grid}}\sum_{i=1}^{n_{grid}}E(\widehat{g}(z_{i})-E\widehat{g}(z_{i}))^{2} \right ]^{\frac{1}{2}}$,
and  $RMSE= \sqrt{bias^{2}+SD^{2}}$, where $z_{i}(i=1,...,n_{grid})$ are grid points on which $g(\cdot)$ is evaluated,  which firstly randomly generated from the distribution of  $\bX$ and then  fixed,  $n_{grid}=2048$ is the number of  grid points, and $E\widehat{g}(z_{i})$ is approximated by its sample mean based on $100$ replications.

\subsection{Data generation}

Denote $\bX_i=(X_{i1},\cdots,X_{id})^\top$ with  each component of $\bX_{i}$ being  {\it {i.i.d.}} generated from a uniform distribution $U(0,1)$.  We consider three target functions: (1) $g_{5}(\bX_i)=x_{i1}^{3}+x_{i2}^{2}+x_{i3}+\left | x_{i4} \right |+cos(x_{i5})$; (2)
$g_{10}(\bX_i)=x_{i1}^{3}+x_{i2}^{2}+x_{i3}+\left | x_{i4} \right |+cos(x_{i5})+sin(x_{i6})+e^{x_{i7}}+log(1+x_{i8})+x_{i9}^{{\frac{1}{2}}}+x_{i10}^{{\frac{1}{3}}}$; and (3) $g_{20}(\bX_{i})=x_{i1}^{5}+x_{i2}^{4}+x_{i3}^{3}+x_{i4}^{2}+x_{i5}+\left | x_{i6} \right |+x_{i7}^{{\frac{1}{2}}}+x_{i8}^{{\frac{1}{3}}}+x_{i9}^{{\frac{1}{4}}}+x_{i10}^{{\frac{1}{5}}}+\left | x_{i11}^{3} \right |+cos(x_{i12})+sin(x_{i13})+cos(x_{i14}^{2})+sin(x_{i15}^{2})+e^{x_{i16}}+log(1+x_{i17})+e^{x_{i18}^{2}}+log(1+x_{i19}^{2})+log(1+x_{i20}^{{\frac{1}{2}}})$,
which are  $p=5, 10$ and 20-dimensional functions, respectively,
where $x_{ij}= \bX_{i}^{\top}\bbeta_{j}, j=1,..., 20$,   $\bbeta_j$ is a $d$-dimensional vector with  $\bbeta_j\left[\left((j-1) \times \left \lfloor \frac{d}{p} \right \rfloor+1 \right) :
\left( j \times \left \lfloor \frac{d}{p} \right \rfloor \right)\right] = \frac{\left(\bgamma^\top, \cdots, \bgamma^\top\right)}{\left \lfloor \frac{d}{20p} \right \rfloor \times \|\bgamma\|_1 }$  and the remaining components of $\bbeta_j$ are $0$, where $\bgamma = (1, 2, \cdots, 20)^\top$. In a word, the non-zero elements of $\bbeta_j$ are integer values ranging from 1 to 20 but scaled according to $L_1$ norms.


We consider the following four distributions for the error $\epsilon_i$:
(I) Standard normal distribution: $\epsilon_i \sim  \mathcal{N}(0,1)$; (II) Mixture gaussion distribution: $\epsilon_i \sim 0.7 \mathcal{N}(0,1)+ 0.3 \mathcal{N}(0,5)$;
(III) Student-t distribution: $\epsilon_i \sim  t(2)$; and (IV) Heteroscedasticity: $\epsilon_{i} \sim  \mathcal{N}(0,3X_{i1}+4X_{i2})$.

We then generate $Y_i$ by $Y_{i} = g_{p}(\bX_{i})+\varepsilon_{i}$  with $p = 5, 10, 20$, respectively.  We set $n=256, 1024$  and consider $d=100,200,400,500,600,800$ for  the above three target functions and four error distributions, respectively. 
All simulation results with 100 replications  are presented in Figures \ref{Fig.1} to  \ref{Fig.3}.

\subsection{Numerical Results}

\begin{figure}[H]
\centering
\includegraphics[width=6in,height=4in]{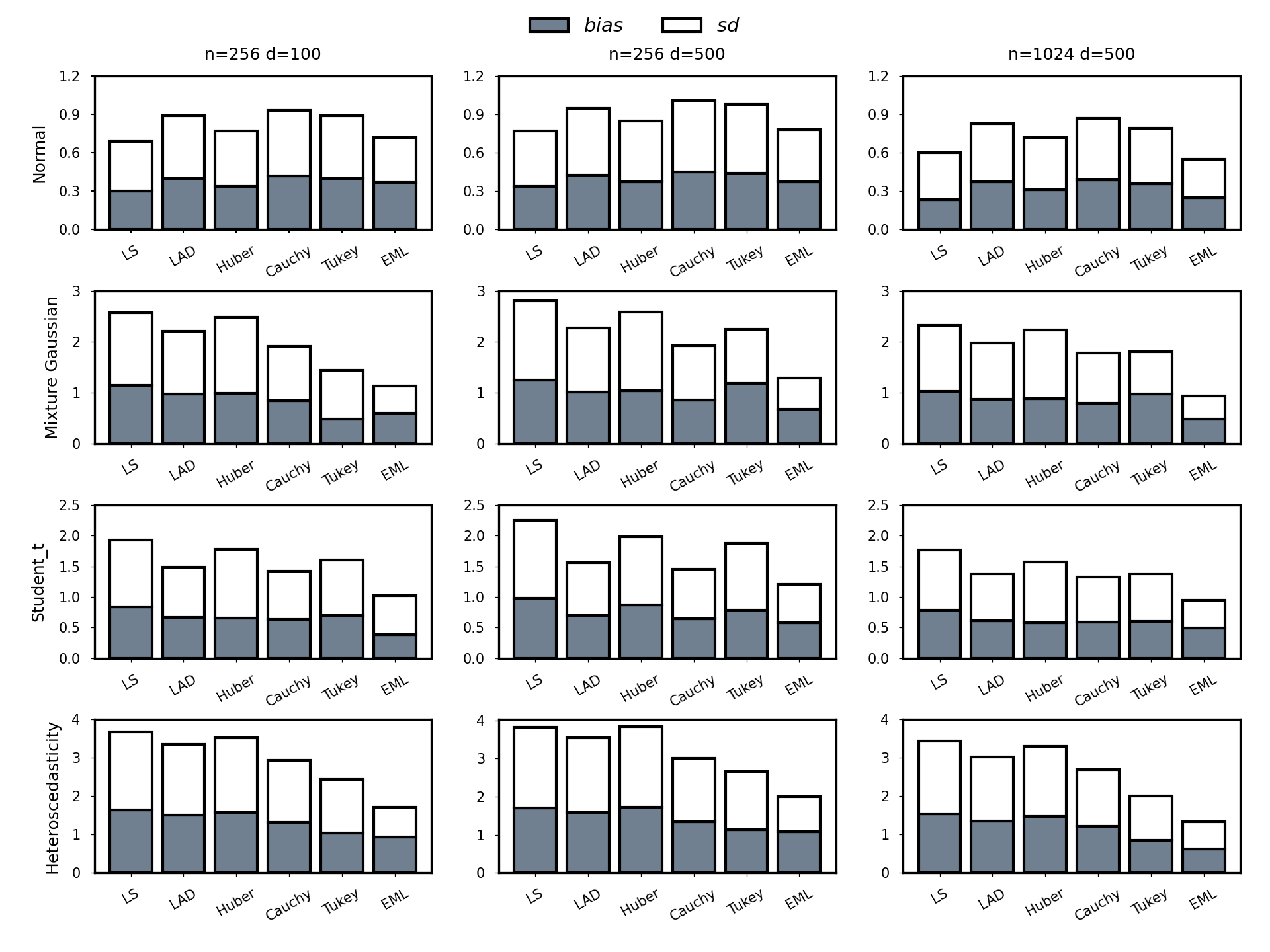}
\caption{The bar chart of the bias and standard deviation of $g_{5}$ using six  methods for four error distributions with sample sizes $n=256, 1024$ and input dimensions $d=100, 500$, respectively.}
\label{Fig.1}
\end{figure}

\begin{figure}[H]
\centering
\includegraphics[width=1\textwidth]{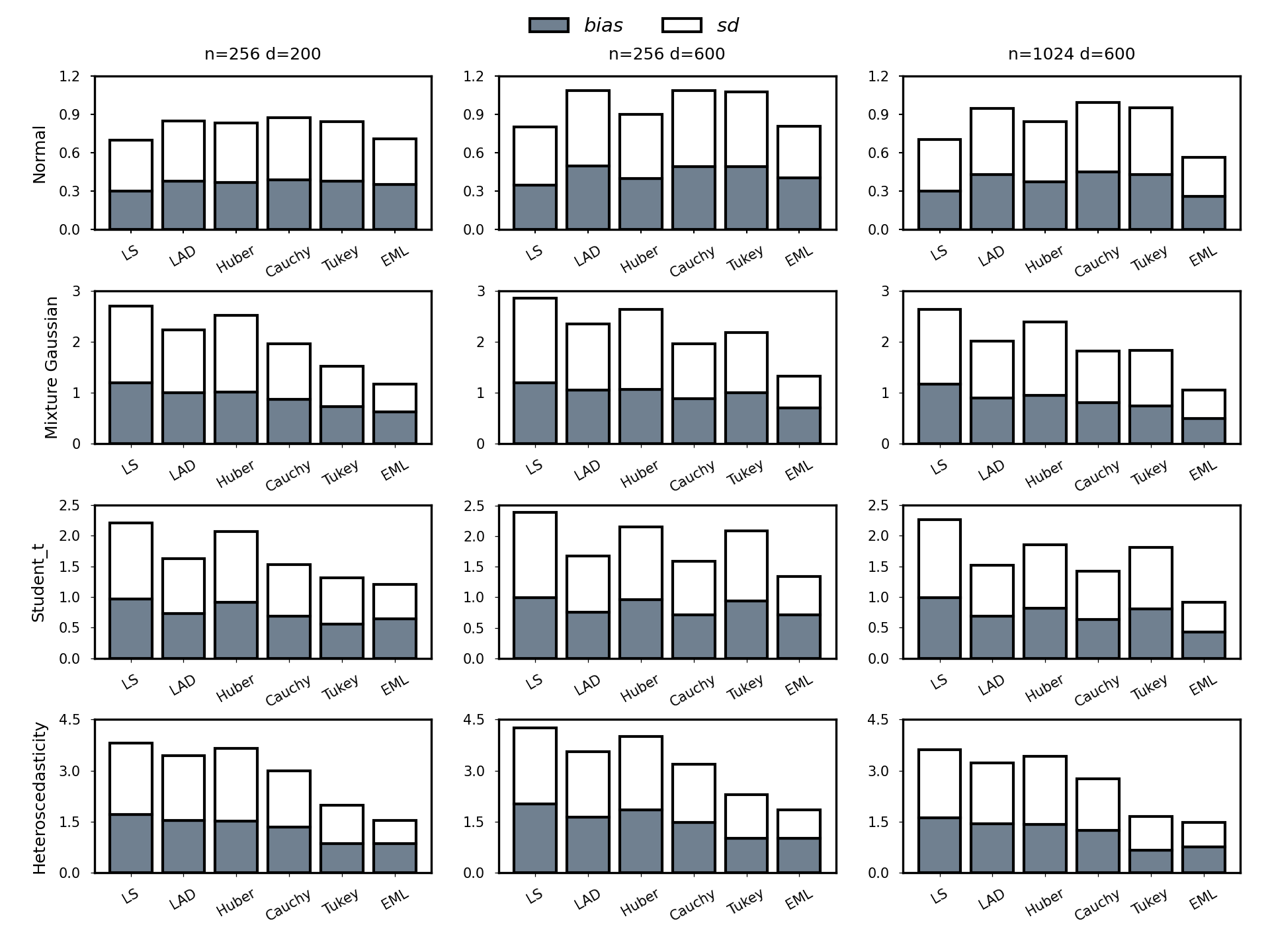}
\caption{The bar chart of the bias and standard deviation of $g_{10}$ using six  methods for four error distributions with sample sizes $n=256, 1024$ and input dimensions $d=200, 600$, respectively.}
\label{Fig.2}
\end{figure}

\begin{figure}[H]
\centering
\includegraphics[width=1\textwidth]{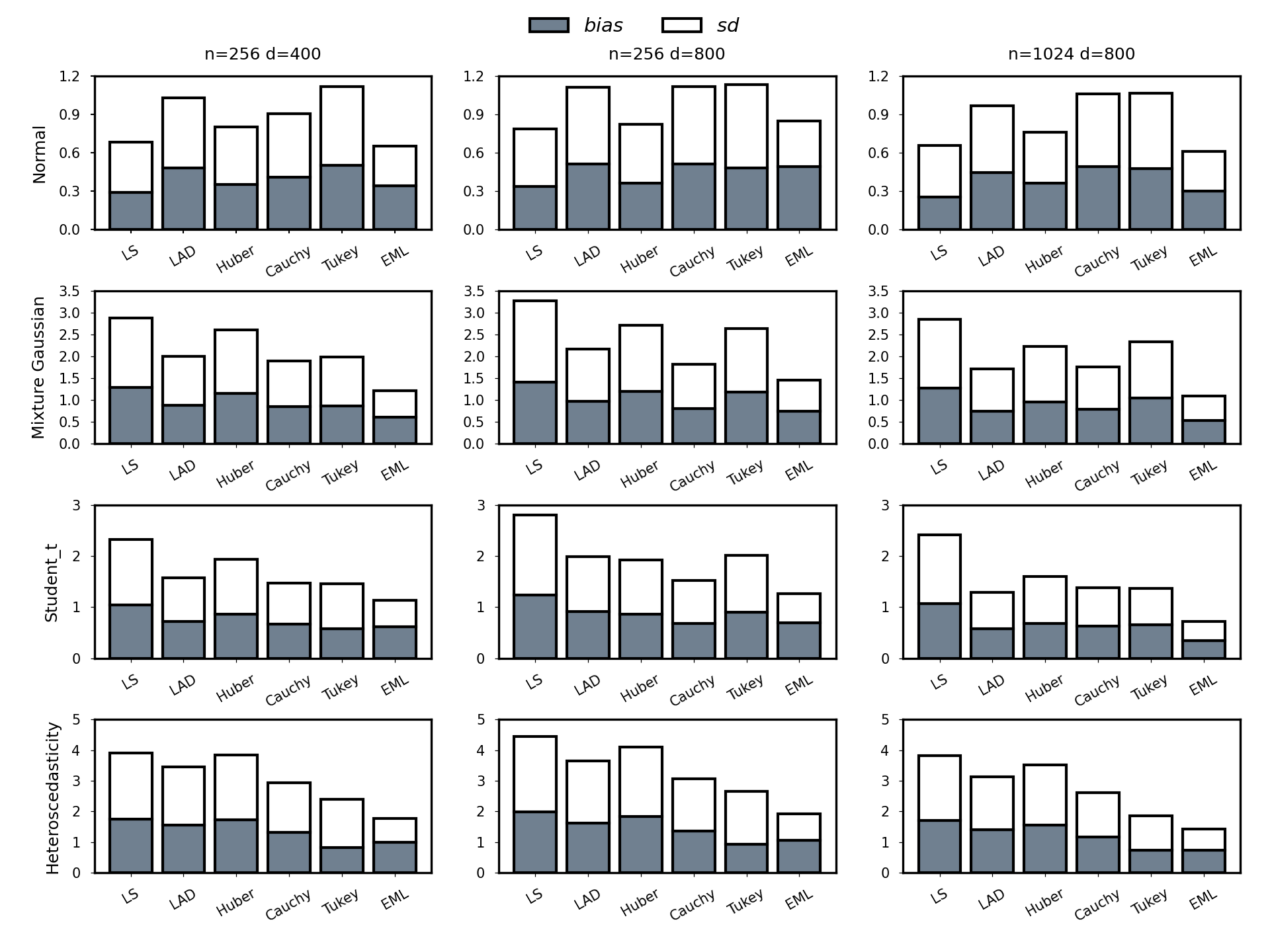}
\caption{The bar chart of the bias and standard deviation of $g_{20}$ using six  methods for four error distributions with sample sizes $n=256, 1024$ and input dimensions $d=400, 800$, respectively.}
\label{Fig.3}
\end{figure}

From Figures \ref{Fig.1} to \ref{Fig.3}, it is evident that the proposed EML consistently and significantly outperforms the robust-based methods in terms of bias, SD, and RMSE. When the errors follow the normal distribution, the LSE is  optimal. In this case, the proposed EML performs comparably to LSE, and both outperform the robust-based methods. This indicates that the loss caused by estimating the density function can be ignored, which aligns with the theoretical findings in Theorem \ref{th:error-one}.
Upon closer examination, we can see  that EML even slightly outperforms LSE for normal data as the sample size increases, for instance, when $n=1024$. This observation implies that the ability of the proposed EML to learn the data structure may become more pronounced as the sample size grows.
For non-normal and heteroscedasticity situations, the LSE performs the worst among all the methods and the proposed EML significantly outperforms the LSE.
Figures \ref{Fig.1} to \ref{Fig.3} also show that the performance of all the methods improves with increasing sample sizes or decreasing dimensions. 

\subsection{Effect of the bandwidth}
\label{band}

\begin{figure}[H]
\centering
\includegraphics[height=8cm,width=15cm]{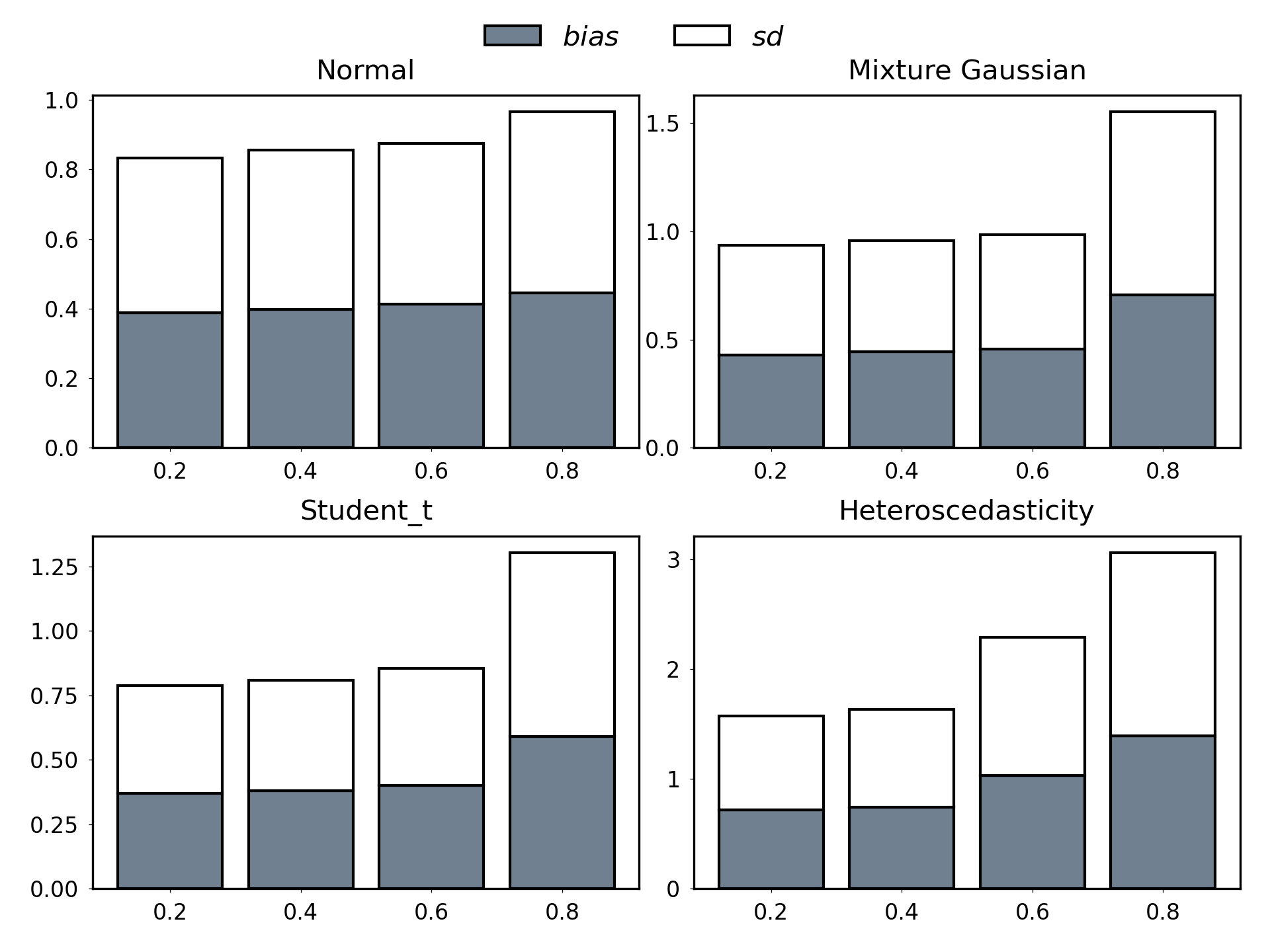}
\caption{The bar chart of the bias and standard deviation of the proposed EML-FNN for $g_5$ under four error distributions with  $(n, d) = (1024,500)$,  as the  bandwidths vary from 0.2 to 0.8.}
\label{Fig.4}
\end{figure}

\begin{figure}[H]
\centering
\includegraphics[height=7cm,width=14cm]{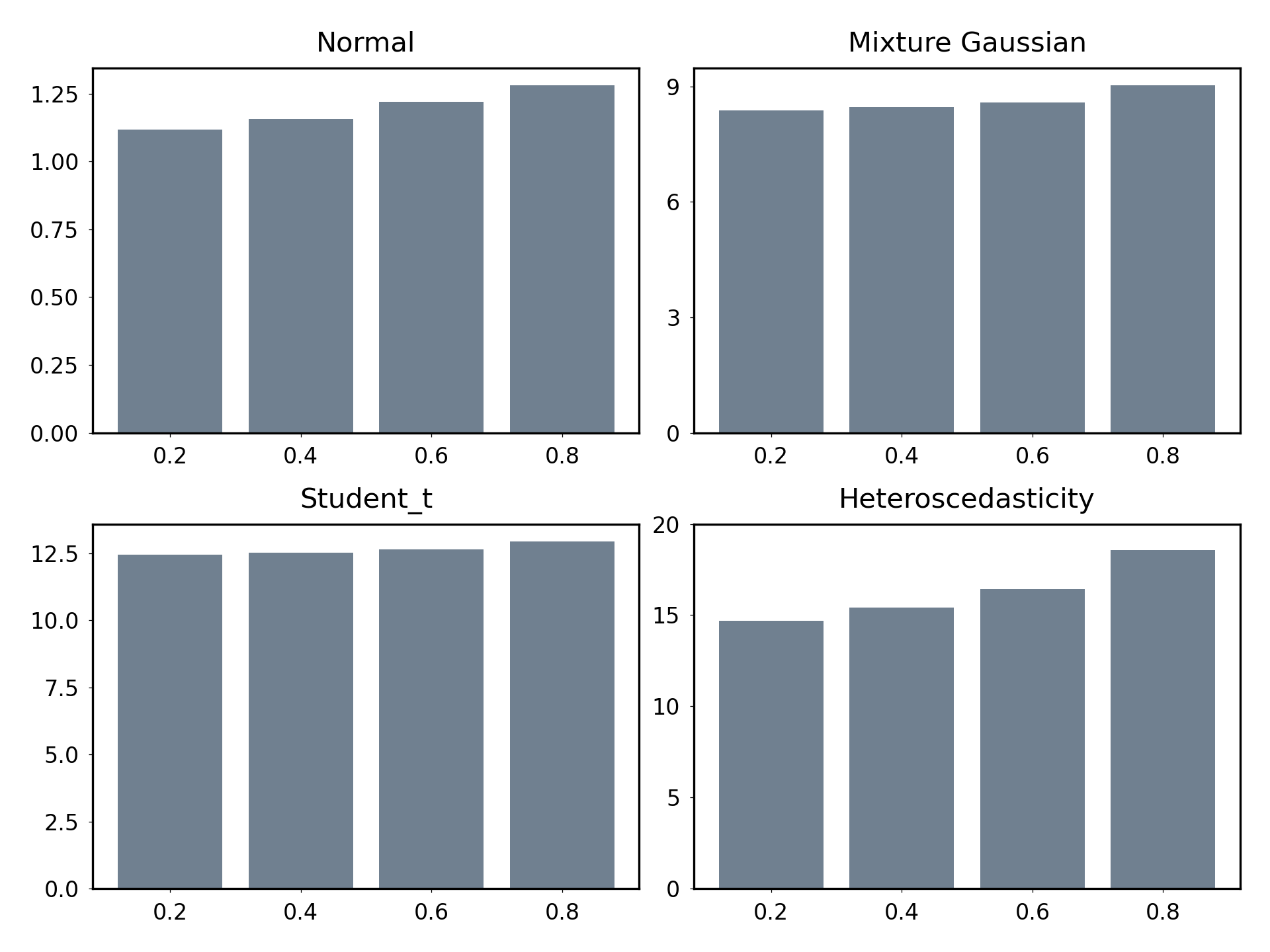}
\caption{The bar chart of the mean prediction error (PE) of the proposed EML-FNN for $g_5$ under four error distributions with  $(n, d) = (1024,500)$,  as the  bandwidths vary from 0.2 to 0.8.}
\label{Fig.5}
\end{figure}

Now, we examine the effect of bandwidth on the proposed method. In Figures \ref{Fig.4} and \ref{Fig.5}, we present the bias, SD, and prediction error (PE) of the EML-FNN estimator when  the bandwidths vary from 0.2 to 0.8 for $g_5$ under four error distributions given  $(n, d) = (1024,500)$.  PE is defined as $PE=\frac{1}{t}\sum_{i=1}^{t}\left \| g(X_{i}^{test})-Y_{i}^{test} \right \|^{2}$, where  $\left \{ (X_{i}^{test},Y_{i}^{test} \right \}_{i=1}^{t}$ represents the test data, which shares the same distribution as the training data.  From  Figures \ref{Fig.4} and \ref{Fig.5}, we can see  that  a smaller bandwidth provides a better estimator in terms of bias, SD, and PE, and the proposed EML estimator is  robust to variations in bandwidth within a certain range that approaches zero.   These  findings are consistent with  the theoretical result presented in Theorem \ref{th:error-one}, which indicates that a small bandwidth is  favored,  and the extra risk is independent of the bandwidth if the bandwidth is appropriately  small. Additionally, the comparison of Figures \ref{Fig.4} and \ref{Fig.5} reveals that the PE is more stable than bias and SD as the bandwidth changes.

\section{Real data example}
\label{realdata}

We applied our proposed EML-FNN and other competing methods to analyze four real  datasets based on the model (\ref{model}) using the observations $(\bX_i, Y_i)_{i=1}^n$.
\begin{itemize}
\item[{(1)}] Boston House Price Dataset.  It is available in the scikit-learn library \citep{scikit-learn} and encompasses a total of $n=506$ observations. The purpose of the analysis is  predicting the house price  based on the  13 input variables $\bX_i$,  such as  urban crime rates, nitric oxide levels, average number of rooms in a dwelling, weighted distance to central areas, and average owner-occupied house prices, among others.
    Following \citet{kong2012single} and \citet{zhou2019efficient},  we employ the logarithm of the median price of owner-occupied residences  in units of \$1,000 as our response  $Y_i$.

\item[{(2)}] QSAR aquatic toxicity Dataset. The dataset was  provided by \citet{cassotti2014prediction} and was used to develop quantitative regression QSAR models for predicting acute aquatic toxicity towards Daphnia Magna. It consists of a total of $n=546$ observations, each has  8 molecular descriptors serving  as covariates $\bX_i$,  including PSA(Tot) (Molecular properties), SAacc (Molecular properties), H-050 (Atom-centred fragments), MLOGP (Molecular properties), RDCHI (Connectivity indices), GATS1p (2D autocorrelations), nN (Constitutional indices), C-040 (Atom-centred fragments).  The response variable $Y_i$ is  the acute aquatic toxicity, specifically the LC50, which is defined as the concentration causing death in 50\% of the test D. magna over a test duration of 48 hours.

\item[{(3)}] QSAR fish toxicity Dataset. Another version of the dataset for quantitative regression QSAR models was provided by \citet{cassotti2015similarity}. This dataset includes 908 observations, each observation has 6 input variables ($\bX_i$) including molecular descriptors: MLOGP (molecular properties), CIC0 (information indices), GATS1i (2D autocorrelations), NdssC (atom-type counts), NdsCH ((atom-type counts), SM1\_Dz(Z) (2D matrix-based descriptors). The response  variable $Y_i$ is the LC50  which is the concentration that causes death in 50\% of test fish over a test duration of 96 hours. 

\item[{(4)}]  Temperature forecast Dataset. The dataset was provided by \citet{cho2020comparative} and aims to correcting bias  of next-day maximum and minimum air temperatures forecast from the LDAPS model operated by the Korea Meteorological Administration over Seoul, South Korea. The data consists of summer data  spanning  from 2013 to 2017.   The input data $\bX_i$ is largely composed of predictions from the LDAPS model for the subsequent day, in-situ records of present-day maximum and minimum temperatures, and  five geographic auxiliary variables. In this dataset, two outputs ($Y_i$) are featured: next-day maximum and minimum air temperatures.

\end{itemize}

\begin{table}
\caption{Mean prediction error for four  real datasets.\label{table:1}}
\resizebox{\textwidth}{60mm}{
\begin{tabular}{c|ccc}

\hline
                                       & \multicolumn{1}{c|}{\textbf{Boston}}                        & \multicolumn{1}{c|}{\textbf{Aquatic Toxicity}}                       & \textbf{Fish Toxicity}                          \\
                                       & \multicolumn{1}{c|}{\citep{scikit-learn}}                   & \multicolumn{1}{c|}{\citep{cassotti2014prediction}}                  & \citep{cassotti2015similarity} \\ \hline
\textbf{LS}                            & \multicolumn{1}{c|}{0.1045}                                 & \multicolumn{1}{c|}{1.2812}                                          & 2.0153                                          \\
\textbf{LAD}                           & \multicolumn{1}{c|}{0.1054}                                 & \multicolumn{1}{c|}{1.2184}                                          & 2.142                                           \\
\textbf{Huber}                         & \multicolumn{1}{c|}{0.1155}                                 & \multicolumn{1}{c|}{1.2003}                                          & 2.1403                                          \\
\textbf{Cauchy}                        & \multicolumn{1}{c|}{ 0.1192}                                & \multicolumn{1}{c|}{ 1.2697}                                         & 2.1179                                          \\
\textbf{Tukey’s biweight}              & \multicolumn{1}{c|}{0.1153}                                 & \multicolumn{1}{c|}{1.3148}                                          & 2.1436                                          \\
\textbf{EML}                           & \multicolumn{1}{c|}{\textbf{0.0833}}                        & \multicolumn{1}{c|}{ \textbf{1.1497}}                                & \textbf{1.8918}                                 \\ \hline
                                       & \textbf{Temperature Forecast}                               &                                                                     &                                                 \\
                                       & \citep{cho2020comparative}                                  &                                                                     &                                \\ \hline
\textbf{LS}                            & 10.9303                                                     & 5.5969                                                               &                                                 \\
\textbf{LAD}                           & 10.6615                                                     & 5.71                                                                 &                                                 \\
\textbf{Huber}                         & 10.6054                                                     & 5.3211                                                               &                                                 \\
\textbf{Cauchy}                        & 11.4592                                                     & 6.0396                                                               &                                 \\
\textbf{Tukey’s biweight}              & 11.2879                                                     & 5.2332                                                               &                                                 \\
\textbf{EML}                           & \textbf{4.4085}                                             & \textbf{2.2196}                                                      &                                 \\ \hline

\end{tabular}}

\end{table}

\begin{figure}[H]
\centering
\includegraphics[width=0.95\textwidth]{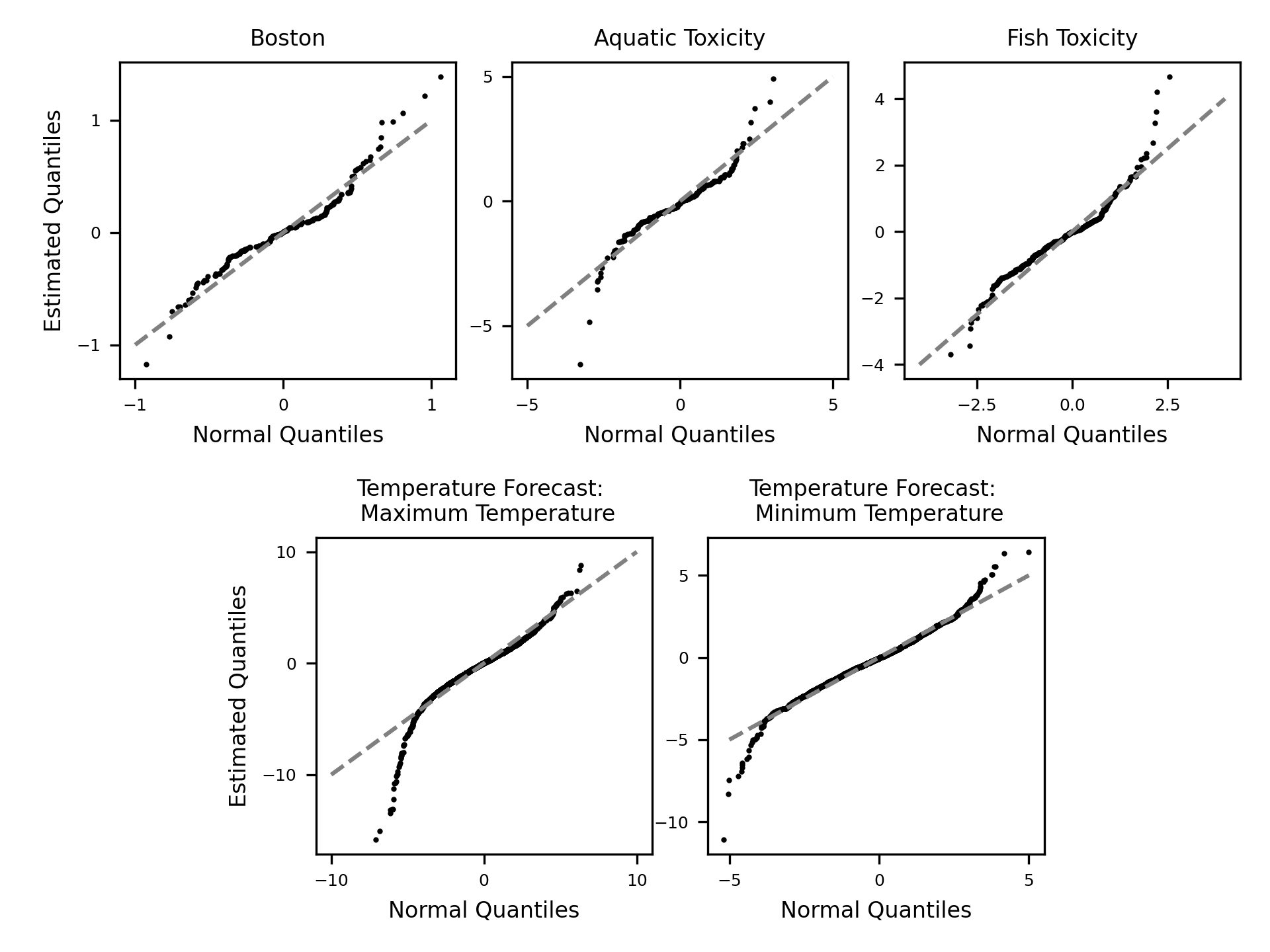}
\caption{The Q-Q plot of the estimated density function for four  real datasets.}
\label{Fig.6}
\end{figure}

We preprocessed all the datasets by applying Z-score normalization to each predictor variable. Inspired from transfer learning, we employed the widely-used fine-tuning technique to simplify the computation.
We initiated the process by training a single network model based on, for example,   the Cauchy loss function  by  employing the methodology outlined in Section \ref{sec:ss}. Subsequently, we leveraged this trained model as a foundation to train all other models with a learning rate of $0.00003$.
All four datasets were randomly split into training and test sets with a ratio of 4:1  to calculate PE.  The entire procedure  was repeated 50 times, and  the average PE were calculated and  presented in Table \ref{table:1}.  The results in Table \ref{table:1} clearly  demonstrate the significant superiority of our approach over other competing methods in terms of a remarkable improvement in prediction accuracy across all four datasets.  Particularly noteworthy is the outstanding performance achieved when applying our proposed EML technique to the Temperature Forecast dataset, where  the improvement of the prediction accuracy achieves up to $50\%$.  To understand the underlying reasons behind this improvement, we proceeded to plot  the Q-Q plot in Figure \ref{Fig.6} 
on the estimated error distribution for each of  four  real datasets.  From the Q-Q plot in Figure \ref{Fig.6}, 
we can see  the distribution of Boston house prices quite close to  normal distribution.  Following this, the toxicity data exhibits  relatively closer resemblance to normality, characterized by a limited number of outliers. In contrast, the temperature data diverges substantially from the normal distribution, with a notable prevalence of extreme values. 
Based on these findings, we can draw the conclusion that the prediction performances, as illustrated in Table \ref{table:1}, are linked  to the degree to which the respective distributions adhere to normality.  Furthermore, from Tables \ref{table:1} and  Figure \ref{Fig.6}, we also can see  that  all the methods exhibit enhanced predictive accuracy when handling datasets that  are more similar to a normal distribution.
This observation further highlights the influence of distribution characteristics on the resulting estimator and  emphasizes the importance  of incorporating distribution information into the analysis.

\section{Concluding Remarks}
\label{conc}

The paper presents an innovative approach to nonparametric regression using FNN. This approach is characterized by its efficiency in both estimation and computation, its adaptability to diverse data distributions, and its robustness in the presence of noise and uncertainty. The key contributions are as follows:  (1) Estimation efficiency:  The method introduces a novel loss function that incorporates not only observed data but also potentially implicit information about the  data structure. By integrating this hidden information, the loss function transforms into an estimated maximum likelihood function, resulting in
desirable properties such as  efficiency. (2) Distribution-free:  The method is independent of data distribution assumptions. Consequently, the approach adeptly handles data with varying distributional characteristics, such as  heavy tails, multimodal distributions, and heterogeneity. (3) Probabilistic Robustness:   The loss function is formulated through a probabilistic framework. This probabilistic approach effectively  reduce the impact of substantial noise and outliers within the data, thereby enhancing its robustness.
(4) Kernel-Based Smoothness:  The method leverages the inherent smoothness  of kernel functions. This enables the calculation of gradients and addresses challenges related to non-differentiability, when dealing with densities such as uniform distributions, mixture distributions, and cases of heteroscedasticity. (5) Computational efficiency: The proposed loss function exclusively involves the regression function $g$. This design  facilitates the straightforward utilization of existing software packages, simplifying the computational and programming.
In summary, the method's capacity to accommodate various data distributions without the need for distributional assumptions renders it versatile and applicable to a wide range of real-world scenarios.

By utilizing a reasonably small bandwidth, 
the proposed estimator is proved to be equivalent to the maximum likelihood estimator (MLE) where the density function is known.  Furthermore, it nearly attains the minimax optimal rate, with only an additional logarithmic factor. Its exceptional performance is further exemplified through comprehensive simulation studies and its successful application to four distinct real-world datasets.

There are several directions for future research. First, it might be possible to extend our method to a more complicated model, such as  the generalized regression  model for a discrete response.  Secondly, practical scenarios often involve multiple responses that exhibit correlations, as seen in  the Temperature Forecast Dataset's maximum and minimum air temperatures. By further modeling  inter-response correlations, predictive capabilities could be enhanced. Lastly, it remains our responsibility to consistently enhance the associated software packages, ensuring seamless application. Despite having introduced an efficient and user-friendly package named EML-FNN, continued optimization and refinement are necessary.

\normalem
\bibliographystyle{apalike}
\bibliography{mybib}

\begin{thebibliography}{}

\bibitem[Bartlett et~al., 2019]{bartlett2019nearly}
Bartlett, P.~L., Harvey, N., Liaw, C., and Mehrabian, A. (2019).
\newblock Nearly-tight vc-dimension and pseudodimension bounds for piecewise
  linear neural networks.
\newblock {\em The Journal of Machine Learning Research}, 20(1):2285--2301.

\bibitem[Bassett~Jr and Koenker, 1978]{bassett1978asymptotic}
Bassett~Jr, G. and Koenker, R. (1978).
\newblock Asymptotic theory of least absolute error regression.
\newblock {\em Journal of the American Statistical Association},
  73(363):618--622.

\bibitem[Bauer and Kohler, 2019]{bauer2019deep}
Bauer, B. and Kohler, M. (2019).
\newblock On deep learning as a remedy for the curse of dimensionality in
  nonparametric regression.
\newblock {\em The Annals of Statistics}, 47(4):2261--2285.

\bibitem[Beaton and Tukey, 1974]{beaton1974fitting}
Beaton, A.~E. and Tukey, J.~W. (1974).
\newblock The fitting of power series, meaning polynomials, illustrated on
  band-spectroscopic data.
\newblock {\em Technometrics}, 16(2):147--185.

\bibitem[Berlinet and Thomas-Agnan, 2011]{berlinet2011reproducing}
Berlinet, A. and Thomas-Agnan, C. (2011).
\newblock {\em Reproducing kernel Hilbert spaces in probability and
  statistics}.
\newblock Springer Science \& Business Media.

\bibitem[Bhattacharya et~al., 2023]{bhattacharya2023deep}
Bhattacharya, S., Fan, J., and Mukherjee, D. (2023).
\newblock Deep neural networks for nonparametric interaction models with
  diverging dimension.
\newblock {\em arXiv preprint arXiv:2302.05851}.

\bibitem[Breiman, 2001]{breiman2001random}
Breiman, L. (2001).
\newblock Random forests.
\newblock {\em Machine learning}, 45:5--32.

\bibitem[Breiman, 2017]{breiman2017classification}
Breiman, L. (2017).
\newblock {\em Classification and regression trees}.
\newblock Routledge.

\bibitem[Cassotti et~al., 2014]{cassotti2014prediction}
Cassotti, M., Ballabio, D., Consonni, V., Mauri, A., Tetko, I.~V., and
  Todeschini, R. (2014).
\newblock Prediction of acute aquatic toxicity toward daphnia magna by using
  the ga-k nn method.
\newblock {\em Alternatives to Laboratory Animals}, 42(1):31--41.

\bibitem[Cassotti et~al., 2015]{cassotti2015similarity}
Cassotti, M., Ballabio, D., Todeschini, R., and Consonni, V. (2015).
\newblock A similarity-based qsar model for predicting acute toxicity towards
  the fathead minnow (pimephales promelas).
\newblock {\em SAR and QSAR in Environmental Research}, 26(3):217--243.

\bibitem[Chen et~al., 2019]{chen2019nonparametric}
Chen, M., Jiang, H., Liao, W., and Zhao, T. (2019).
\newblock Nonparametric regression on low-dimensional manifolds using deep relu
  networks: Function approximation and statistical recovery.
\newblock {\em arXiv preprint arXiv:1908.01842}.

\bibitem[Cheng, 1984]{cheng1984strong}
Cheng, P.~E. (1984).
\newblock Strong consistency of nearest neighbor regression function
  estimators.
\newblock {\em Journal of Multivariate Analysis}, 15(1):63--72.

\bibitem[Cho et~al., 2020]{cho2020comparative}
Cho, D., Yoo, C., Im, J., and Cha, D.-H. (2020).
\newblock Comparative assessment of various machine learning-based bias
  correction methods for numerical weather prediction model forecasts of
  extreme air temperatures in urban areas.
\newblock {\em Earth and Space Science}, 7(4):e2019EA000740.

\bibitem[Cleveland, 1979]{cleveland1979robust}
Cleveland, W.~S. (1979).
\newblock Robust locally weighted regression and smoothing scatterplots.
\newblock {\em Journal of the American statistical association},
  74(368):829--836.

\bibitem[Devroye et~al., 1994]{devroye1994strong}
Devroye, L., Gyorfi, L., Krzyzak, A., and Lugosi, G. (1994).
\newblock On the strong universal consistency of nearest neighbor regression
  function estimates.
\newblock {\em The Annals of Statistics}, 22(3):1371--1385.

\bibitem[Donoho et~al., 1995]{donoho1995wavelet}
Donoho, D.~L., Johnstone, I.~M., Kerkyacharian, G., and Picard, D. (1995).
\newblock Wavelet shrinkage: asymptopia?
\newblock {\em Journal of the Royal Statistical Society: Series B
  (Methodological)}, 57(2):301--337.

\bibitem[Fan and Gijbels, 2018]{fan2018local}
Fan, J. and Gijbels, I. (2018).
\newblock {\em Local polynomial modelling and its applications}.
\newblock Routledge.

\bibitem[Fan and Gu, 2022]{fan2022factor}
Fan, J. and Gu, Y. (2022).
\newblock Factor augmented sparse throughput deep relu neural networks for high
  dimensional regression.
\newblock {\em arXiv preprint arXiv:2210.02002}.

\bibitem[Fan et~al., 2022]{fan2022noise}
Fan, J., Gu, Y., and Zhou, W.-X. (2022).
\newblock How do noise tails impact on deep relu networks?
\newblock {\em arXiv preprint arXiv:2203.10418}.

\bibitem[Farrell et~al., 2021]{farrell2021deep}
Farrell, M.~H., Liang, T., and Misra, S. (2021).
\newblock Deep neural networks for estimation and inference.
\newblock {\em Econometrica}, 89(1):181--213.

\bibitem[Gy{\"o}rfi et~al., 2002]{gyorfi2002distribution}
Gy{\"o}rfi, L., Kohler, M., Krzyzak, A., Walk, H., et~al. (2002).
\newblock {\em A distribution-free theory of nonparametric regression},
  volume~1.
\newblock Springer.

\bibitem[Hall and Huang, 2001]{hall2001nonparametric}
Hall, P. and Huang, L.-S. (2001).
\newblock Nonparametric kernel regression subject to monotonicity constraints.
\newblock {\em The Annals of Statistics}, 29(3):624--647.

\bibitem[H{\"a}rdle, 1989]{hardle1989asymptotic}
H{\"a}rdle, W. (1989).
\newblock Asymptotic maximal deviation of m-smoothers.
\newblock {\em Journal of Multivariate Analysis}, 29(2):163--179.

\bibitem[He et~al., 2015]{he2015delving}
He, K., Zhang, X., Ren, S., and Sun, J. (2015).
\newblock Delving deep into rectifiers: Surpassing human-level performance on
  imagenet classification.
\newblock In {\em 2015 IEEE International Conference on Computer Vision
  (ICCV)}, pages 1026--1034.

\bibitem[He et~al., 2013]{he2013quantile}
He, X., Wang, L., and Hong, H.~G. (2013).
\newblock Quantile-adaptive model-free variable screening for high-dimensional
  heterogeneous data.
\newblock {\em The Annals of Statistics}, 41(1):342--369.

\bibitem[Huber, 1973]{huber1973robust}
Huber, P.~J. (1973).
\newblock {Robust Regression: Asymptotics, Conjectures and Monte Carlo}.
\newblock {\em The Annals of Statistics}, 1(5):799 -- 821.

\bibitem[Jiao et~al., 2021]{jiao2021deep}
Jiao, Y., Shen, G., Lin, Y., and Huang, J. (2021).
\newblock Deep nonparametric regression on approximately low-dimensional
  manifolds.
\newblock {\em arXiv preprint arXiv:2104.06708}.

\bibitem[Kingma and Ba, 2014]{kingma2014adam}
Kingma, D.~P. and Ba, J. (2014).
\newblock Adam: A method for stochastic optimization.
\newblock {\em arXiv preprint arXiv:1412.6980}.

\bibitem[Koenker and Bassett~Jr, 1978]{koenker1978regression}
Koenker, R. and Bassett~Jr, G. (1978).
\newblock Regression quantiles.
\newblock {\em Econometrica}, 46(1):33--50.

\bibitem[Kohler et~al., 2022]{kohler2022estimation}
Kohler, M., Krzyzak, A., and Langer, S. (2022).
\newblock Estimation of a function of low local dimensionality by deep neural
  networks.
\newblock {\em IEEE transactions on information theory}, 68(6):4032--4042.

\bibitem[Kohler and Langer, 2021]{kohler2021rate}
Kohler, M. and Langer, S. (2021).
\newblock On the rate of convergence of fully connected deep neural network
  regression estimates.
\newblock {\em The Annals of Statistics}, 49(4):2231--2249.

\bibitem[Kong and Xia, 2012]{kong2012single}
Kong, E. and Xia, Y. (2012).
\newblock A single-index quantile regression model and its estimation.
\newblock {\em Econometric Theory}, 28(4):730--768.

\bibitem[Lederer, 2020]{lederer2020risk}
Lederer, J. (2020).
\newblock Risk bounds for robust deep learning.
\newblock {\em arXiv preprint arXiv:2009.06202}.

\bibitem[Liu et~al., 2022]{liu2022optimal}
Liu, R., Boukai, B., and Shang, Z. (2022).
\newblock Optimal nonparametric inference via deep neural network.
\newblock {\em Journal of Mathematical Analysis and Applications},
  505(2):125561.

\bibitem[Loftsgaarden and Quesenberry, 1965]{loftsgaarden1965nonparametric}
Loftsgaarden, D.~O. and Quesenberry, C.~P. (1965).
\newblock A nonparametric estimate of a multivariate density function.
\newblock {\em The Annals of Mathematical Statistics}, 36(3):1049--1051.

\bibitem[Lu et~al., 2021]{lu2021deep}
Lu, J., Shen, Z., Yang, H., and Zhang, S. (2021).
\newblock Deep network approximation for smooth functions.
\newblock {\em SIAM Journal on Mathematical Analysis}, 53(5):5465--5506.

\bibitem[Lv et~al., 2018]{lv2018oracle}
Lv, S., Lin, H., Lian, H., and Huang, J. (2018).
\newblock Oracle inequalities for sparse additive quantile regression in
  reproducing kernel hilbert space.
\newblock {\em The Annals of Statistics}, 46(2):781--813.

\bibitem[Nadaraya, 1964]{nadaraya1964estimating}
Nadaraya, E.~A. (1964).
\newblock On estimating regression.
\newblock {\em Theory of Probability \& Its Applications}, 9(1):141--142.

\bibitem[Nakada and Imaizumi, 2019]{nakada2019adaptive}
Nakada, R. and Imaizumi, M. (2019).
\newblock Adaptive approximation and estimation of deep neural network to
  intrinsic dimensionality.
\newblock {\em arXiv preprint arXiv:1907.02177}, 6.

\bibitem[Onzon, 2011]{onzon2011multivariate}
Onzon, E. (2011).
\newblock Multivariate cram{\'e}r--rao inequality for prediction and efficient
  predictors.
\newblock {\em Statistics \& probability letters}, 81(3):429--437.

\bibitem[Padilla et~al., 2020]{padilla2020quantile}
Padilla, O. H.~M., Tansey, W., and Chen, Y. (2020).
\newblock Quantile regression with deep relu networks: Estimators and minimax
  rates.
\newblock {\em arXiv preprint arXiv:2010.08236}.

\bibitem[Paszke et~al., 2019]{paszke2019pytorch}
Paszke, A., Gross, S., Massa, F., Lerer, A., Bradbury, J., Chanan, G., Killeen,
  T., Lin, Z., Gimelshein, N., Antiga, L., et~al. (2019).
\newblock Pytorch: An imperative style, high-performance deep learning library.
\newblock {\em Advances in neural information processing systems}, 32.

\bibitem[Pedregosa et~al., 2011]{scikit-learn}
Pedregosa, F., Varoquaux, G., Gramfort, A., Michel, V., Thirion, B., Grisel,
  O., Blondel, M., Prettenhofer, P., Weiss, R., Dubourg, V., Vanderplas, J.,
  Passos, A., Cournapeau, D., Brucher, M., Perrot, M., and Duchesnay, E.
  (2011).
\newblock Scikit-learn: Machine learning in {P}ython.
\newblock {\em Journal of Machine Learning Research}, 12:2825--2830.

\bibitem[Schmidt-Hieber, 2019]{schmidt2019deep}
Schmidt-Hieber, J. (2019).
\newblock Deep relu network approximation of functions on a manifold.
\newblock {\em arXiv preprint arXiv:1908.00695}.

\bibitem[Schmidt-Hieber, 2020]{schmidt2020nonparametric}
Schmidt-Hieber, J. (2020).
\newblock Nonparametric regression using deep neural networks with relu
  activation function.
\newblock {\em The Annals of Statistics}, 48(4):1875--1897.

\bibitem[Schumaker, 2007]{schumaker2007spline}
Schumaker, L. (2007).
\newblock {\em Spline functions: basic theory}.
\newblock Cambridge university press.

\bibitem[Shen et~al., 2021a]{shen2021deep}
Shen, G., Jiao, Y., Lin, Y., Horowitz, J.~L., and Huang, J. (2021a).
\newblock Deep quantile regression: Mitigating the curse of dimensionality
  through composition.
\newblock {\em arXiv preprint arXiv:2107.04907}.

\bibitem[Shen et~al., 2021b]{shen2021robust}
Shen, G., Jiao, Y., Lin, Y., and Huang, J. (2021b).
\newblock Robust nonparametric regression with deep neural networks.
\newblock {\em arXiv preprint arXiv:2107.10343}.

\bibitem[Shen et~al., 2019]{shen2019deep}
Shen, Z., Yang, H., and Zhang, S. (2019).
\newblock Deep network approximation characterized by number of neurons.
\newblock {\em arXiv preprint arXiv:1906.05497}.

\bibitem[Stone, 1977]{stone1977consistent}
Stone, C.~J. (1977).
\newblock Consistent nonparametric regression1.
\newblock {\em The Annals of Statistics}, 5(4):595--620.

\bibitem[Stone, 1982]{stone1982optimal}
Stone, C.~J. (1982).
\newblock Optimal global rates of convergence for nonparametric regression.
\newblock {\em Ann. Statist.}, 10(1):1040--1053.

\bibitem[Tukey et~al., 1977]{tukey1977exploratory}
Tukey, J.~W. et~al. (1977).
\newblock {\em Exploratory data analysis}, volume~2.
\newblock Reading, MA.

\bibitem[Wang and Scott, 1994]{wang19941}
Wang, F.~T. and Scott, D.~W. (1994).
\newblock The l 1 method for robust nonparametric regression.
\newblock {\em Journal of the American Statistical Association},
  89(425):65--76.

\bibitem[Watson, 1964]{watson1964smooth}
Watson, G.~S. (1964).
\newblock Smooth regression analysis.
\newblock {\em Sankya, Series A}, 26:359--372.

\bibitem[Yan and Yao, 2023]{yan2023nonparametric}
Yan, S. and Yao, F. (2023).
\newblock Nonparametric regression for repeated measurements with deep neural
  networks.
\newblock {\em arXiv preprint arXiv:2302.13908}.

\bibitem[Yarotsky, 2017]{yarotsky2017error}
Yarotsky, D. (2017).
\newblock Error bounds for approximations with deep relu networks.
\newblock {\em Neural Networks}, 94:103--114.

\bibitem[Zhou et~al., 2019]{zhou2019efficient}
Zhou, L., Lin, H., Chen, K., and Liang, H. (2019).
\newblock Efficient estimation and computation of parameters and nonparametric
  functions in generalized semi/non-parametric regression models.
\newblock {\em Journal of Econometrics}, 213(2):593--607.

\bibitem[Zhou et~al., 2018]{zhou2018efficient}
Zhou, L., Lin, H., and Liang, H. (2018).
\newblock Efficient estimation of the nonparametric mean and covariance
  functions for longitudinal and sparse functional data.
\newblock {\em Journal of the American Statistical Association},
  113(524):1550--1564.

\end{thebibliography}

\end{sloppypar}
\end{document}